# Bi-Phasic Quasistatic Brain Communication for Fully Untethered Connected Brain Implants

Baibhab Chatterjee*, Mayukh Nath, Gaurav Kumar K, Shulan Xiao,
Krishna Jayant & Shreyas Sen

**ABSTRACT**

Wireless communication using electro-magnetic (EM) fields acts as the backbone for information exchange among wearable devices around the human body. However, for Implanted devices, EM fields incur high amount of absorption in the tissue, while alternative modes of transmission including ultrasound, optical and magneto-electric methods result in large amount of transduction losses due to conversion of one form of energy to another, thereby increasing the overall end-to-end energy loss. To solve the challenge of wireless *powering and communication in a brain implant with low end-end channel loss*, we present Bi-Phasic Quasistatic Brain Communication (BP-QBC), achieving < 60dB worst-case end-to-end channel loss at a channel length of ~55mm, by using Electro-quasistatic (EQS) Signaling that *avoids transduction losses due to no field-modality conversion*. BP-QBC utilizes dipole coupling based signal transmission within the brain tissue using differential excitation in the transmitter (TX) and differential signal pick-up at the receiver (RX), while offering ~41X lower power w.r.t. traditional Galvanic Human Body Communication (G-HBC) at a carrier frequency of 1MHz, by blocking any DC current paths through the brain tissue. Since the electrical signal transfer through the human tissue is electro-quasistatic up to several 10's of MHz range, BP-QBC allows a scalable (bps-10Mbps) duty-cycled uplink (UL) from the implant to an external wearable. The power consumption in the BP-QBC TX is only 0.52 µW at 1Mbps (with 1% duty cycling), which is within the range of harvested power in the downlink (DL) from a wearable hub to an implant through the EQS brain channel, with externally applied electric currents < 1/5th of ICNIRP safety limits. Furthermore, BP-QBC *eliminates the need for sub-cranial interrogators/repeaters*, as it offers better signal strength due to no field transduction. Such low end-to-end channel loss with high data rates enabled by a completely new modality of brain communication and powering has deep societal and scientific impact in the fields of neurobiological research, brain-machine interfaces, electroceuticals and connected healthcare.

## Introduction

Recent advances in neurobiological research have created considerable demand in implantable brain-machine interfaces (BMIc) with applications in (1) Neural Recording systems for motor/behavioral prediction, and (2) Neurostimulators for clinical therapy/monitoring cell physiology. Future advancements of societally critical applications such as neuroscientific studies, brain-machine interfaces, electroceuticals and connected healthcare would rely on extremely small form-factor implantables[1,2] and/or injectable devices, placed within the central and peripheral nervous systems of *freely moving subjects*, triggering the need for self-sustained, energy-efficient and secure mechanisms for information exchange[3,4].

As shown in Fig. 1a, conventional neural interfaces utilize tethered communication for data transmission and powering[5]. Such wired connections increase the risks of cortical scarring, gliosis, infection, and leakage of cerebrospinal fluid (CSF). As a result, there has been a considerable amount of research effort in the past decade to make these interfaces wireless. Due to the high amount of absorption of traditional EM signals in the brain tissue, alternative modalities such as Optical and Ultrasound signaling have been explored, which in turn incur large transduction losses, and hence require an interrogator/repeater to be surgically placed under the skull to improve the quality of the signal (Fig. 1b). Future of such brain implants is envisioned[6] to consist of a *network of untethered multi-channel nodes, utilizing wireless communication and powering without a sub-cranial interrogator*, as shown in Fig. 1c. This article focuses on a newly developed interrogator-less, wireless communication modality for brain implants, named Bi-Phasic Quasistatic Brain Communication (BP-QBC).

Fig. 2a-c show the state-of-the-art tethered and untethered miniaturized wireless neural sensors[7-9] and stimulators[10-12] that have been demonstrated with various data/power transmission modalities. Radio Frequency (RF)[7,13] suffers from increased tissue absorption at high frequencies, and requires large TX power (0.5 W in previous work[7,13], exceeding ICNIRP safety guidelines[14-16] by ~10X). While Optical (OP)[8] and Ultrasonic (US)[9] telemetry are safer, they suffer from significant loss due to scattering and skull absorption (110 dB loss in earlier work[9]) that reduces end-to-end efficiency, necessitating a sub-cranial interrogator which is surgically placed. The fairly recent Magneto-Electric (ME)[10] technique has low tissue-

    

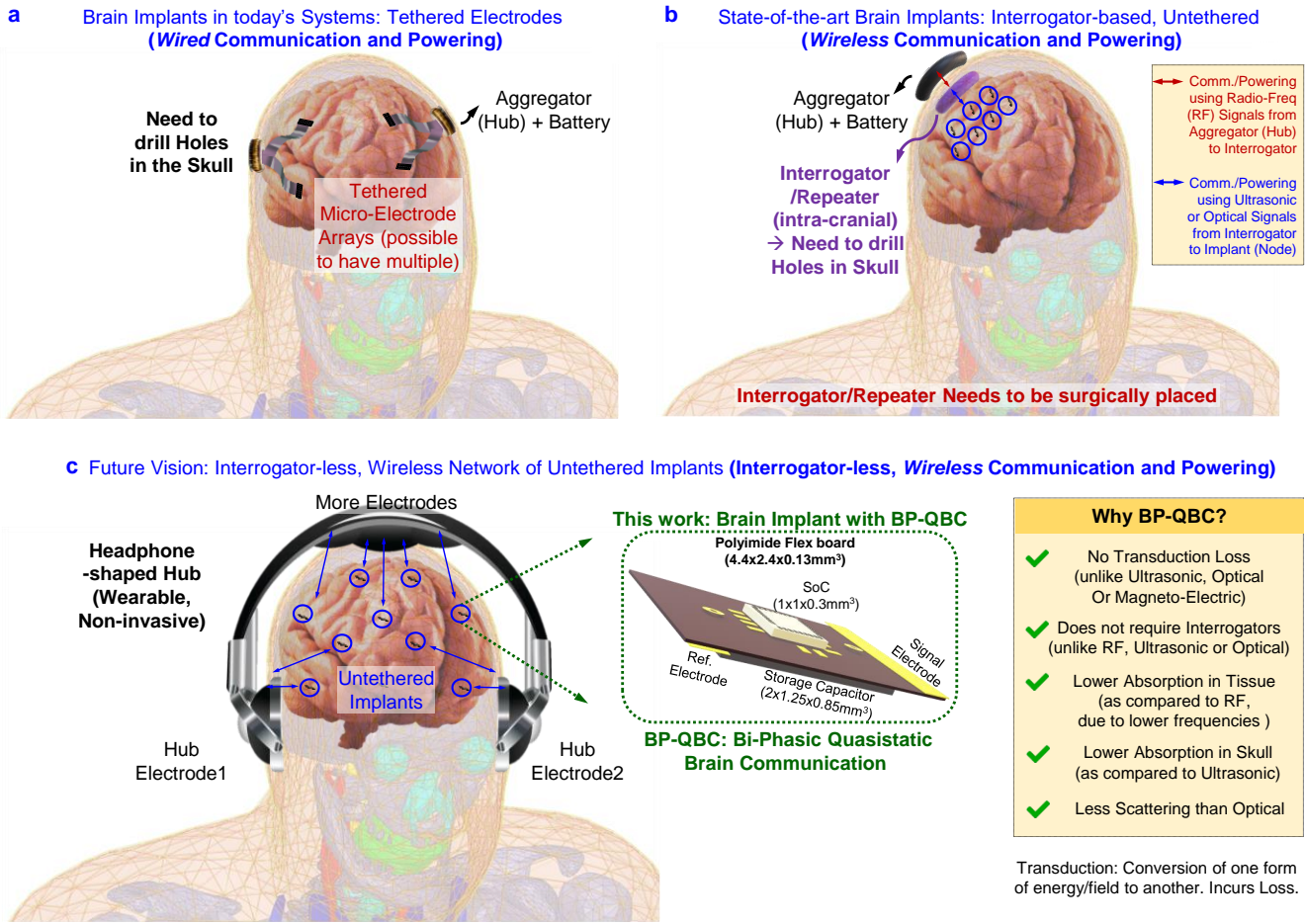

**Fig. 1 | Need for Wireless Communication in Brain Implants. a,** Wired/Tethered Communication and Powering in today's Brain Implants (an example is taken from Neuralink V2.0[5]) wherein the tethered electrodes are placed inside the brain by drilling holes on the skull. The data aggregator and battery are placed in the hole created on the skull. **b,** State-of-the-art brain implants with interrogator/repeaters surgically placed in the sub-cranial region that communicate wirelessly with the implanted nodes using optical/ultrasound techniques. **c,** Untethered Bi-Phasic Quasistatic Brain Communication (BP-QBC) for data communication and powering in grain-sized nodes, sprinkled throughout the brain, which communicate with a headphone-shaped wearable hub. The Vision for Future Brain Implants, with a network of untethered multi-channel implants/nodes, enabled by Wireless Communication and Powering is also shown. *Part of the human figures were created using the human model from* "NEVA EM *LLC*"[46] *and using the open-source software* 'MakeHuman'[49].

absorption but also suffers from high transduction loss (0.1mT magnetic field requirement in previous work[10], which is equivalent to ~300kV/m electric field for iso-energy-density). The pros and cons of the different methods are shown in Fig. 2a-c. As an alternative, Fig. 2b-f also describes BP-QBC for communication in a neural implant. According to the vision in Fig. 1c, the implant can sense and transmit information to a wearable headphone-shaped hub through the UL. The hub sends power and configuration/scan bits to the implant through the DL. Both UL and DL use fully electrical signals to avoid transduction losses (a challenge in OP, US and ME systems). Specifically, the UL utilizes 10s of kHz-10s of MHz narrow-band electro-quasistatic frequencies (with an option of increasing the frequency to ~1 GHz) to (1) avoid interfering with physiological signals and (2) avoid stimulating the brain tissue with low-frequencies. For traditional capacitive HBC, in the ideal scenario (with no geometric/positional imbalance as shown in earlier work[17]), the capacitive return path required for signal transmission is not present separately for the implanted TX to earth's ground, leading to almost zero received voltage at the RX[18]. On the other hand, for galvanic HBC, the electrodes on the implant are almost shorted through the low-impedance (~100s of Ω to a few kΩ) tissue/fluids in the body, resulting in high DC power consumption. In the BP-QBC method, a DC-blocking capacitor in the signal path creates a bi-phasic output for communication that eliminates the DC power going into the tissue resistance and maintains ion balance in the channel. *While bi-phasic signals have been commonly used for neural stimulation, this is the first work demonstrating the bi-phasic modality for communicating with a deep implant.*

## Electro-Quasistatic Brain Communication: Fundamentals

BP-QBC uses electro-quasistatic (EQS) transmission through the conductive layers in the brain tissue. At lower frequencies (several 10's of MHz or less), the transmission can quasistatically be approximated to be electrical in nature.

**Electro-Quasistatic (EQS) Data Transmission**

In traditional low-frequency capacitive as well as galvanic HBC, the potential difference created by the magnetic fields is usually ignored since no closed current loops exist at the transmitting or receiving electrodes. In fact, below a certain frequency ($f$) limit, magnetic fields do not contribute during the data transfer, allowing EQS signal transmission through the brain tissue. The relation between the magnitudes of the developed electric field ($\vec{E}$) and the approximation error ($\vec{E}_{error}$) for EQS data transmission[19,20] is given as shown in equation (1):

$$\vec{E} = \vec{E}_{EQS} + \vec{E}_{error}, \quad \frac{E_{error}}{E} = \omega^2 \mu_{tissue} \epsilon_{tissue} r^2 \tag{1}$$

In equation (1), $r$ represents the dimension of the transmit electrode for EQS-HBC ($< 1$ cm), $\epsilon$ and $\mu$ denotes the permittivity and permeability, respectively, of the conductive tissue layer in the brain. The maximum relative permittivity (for the worst case, at very low frequencies such as kHz) of the brain tissue is ~3000[21,22]. The near-field quasistatic approximation ($\vec{E} \approx \vec{E}_{EQS}$) holds good as long as the magnitude of $E_{error} \ll E$, which implies:

$$\omega^2 \mu_{tissue} \epsilon_{tissue} r^2 \ll 1, \epsilon_{tissue} \approx 3000 \epsilon_{air}, \mu_{tissue} \approx \mu_{air}, \tag{2}$$

$$f \ll \frac{1}{2\pi r \sqrt{\mu_{tissue} \epsilon_{tissue}}} \approx 87.11 \text{ MHz}. \tag{3}$$

This means that the intensity of the electromagnetic fields radiated is dominated by the quasistatic nearfield[22], as long as $f \ll 87.11$ MHz, considering $c = 3 \times 10^8$ m/s to be the velocity of propagation of EM waves in air. However, it must be kept in mind that biological tissue is dispersive, and hence the threshold frequency could vary. In this work, unless otherwise stated, we employ transmission frequency of 1 MHz, thereby allowing the EQS field as the dominant mode of signal propagation through the brain, with an insignificant approximation error = $\left(\frac{1 \text{ MHz}}{87.11 \text{ MHz}}\right)^2 \times 100 = 0.013\%$.

**Pros and Cons of Capacitive EQS HBC (C-HBC) for Brain Implants**

In the EQS range, one of the most common HBC modalities for wearable devices utilize capacitive signal transfer mechanism[23-39], wherein the TX excites the human body using a single-ended electrode, and the RX picks up the signal from a different point on the body using another single-ended electrode. Since there is no common reference between the TX and RX, the signal transfer mechanism becomes a function of the return path capacitances between (1) the TX and earth's ground, and (2) the RX and earth's ground. Bio-physical models for C-HBC were developed in our earlier works[18,34] where the effects of these capacitances on the overall channel transfer function were analyzed, which was found to be proportional to the return path capacitances at both the TX and the RX. The effect of moving the TX device within the body (for an implant) was also analyzed[18], which showed that the return path capacitance for the implanted TX becomes almost zero, as the electric fields terminate within the body, and cannot terminate to the earth's ground as in the wearable scenario. Due to the absence of a TX return path capacitance, the RX received voltage becomes zero. This is shown in Fig 2d for a brain implant. However, due to certain geometrical and positional asymmetries in any realistic device[17], some amount of voltage is generally observed at the RX, but the channel loss is usually as high as 80-100 dB. Interestingly, due to the presence of only one electrode (single-ended excitation), there is no DC current path from the signal electrode to the reference electrode in the TX even if the signal is not DC balanced.

**Pros and Cons of Galvanic EQS HBC (G-HBC) for Brain Implants**

On the other hand, galvanic mode of signal transfer [17,40-43] has been shown to work better for implants, wherein the TX uses differential (galvanic) excitation, and the RX picks up the signal differentially. As shown in Fig. 2d, part of the electric fields going from the signal electrode to the reference electrode in the TX is received at the RX electrodes. However, the human tissue (brain, in this scenario) presents itself as a low-resistance load (~1 kΩ or less) between the signal and reference electrodes of the TX[44]. If the TX signal is not DC balanced (which is usually the case for traditional G-HBC for wearable devices), this will result in a significant amount of DC power for the implant.

**Bi-Phasic Quasistatic Brain Communication (BP-QBC)**

*To simultaneously leverage the advantages of C-HBC and G-HBC*, we present BP-QBC, which utilizes differential electrodes at both the TX and the RX for a signal transfer mechanism similar to G-HBC, but also employs a series capacitor before the signal electrode to prevent any DC current flow into the brain tissue, as shown in Fig. 2b. Being fully electro-quasistatic, BP-QBC does not suffer from the transduction losses which competing wireless techniques (OP/US/ME) are affected with. At the same time, unlike RF, BP-QBC signals are not absorbed as much in the brain tissue because of the lower frequencies. Moreover, there is no requirement of large coils as in RF/Inductive techniques. Furthermore, due to the low end-to-end system loss due to no field transduction, there is no need for sub-cranial interrogators/repeaters to boost the signal, as generally observed in OP and US systems, making the proposed BP-QBC technique amenable to the vision of interrogator-less, wireless network of implants as shown in Fig. 1c.

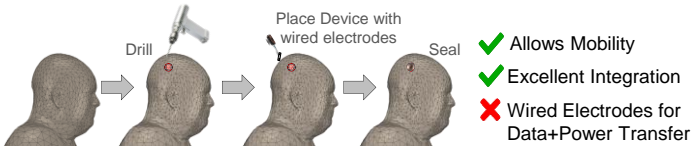
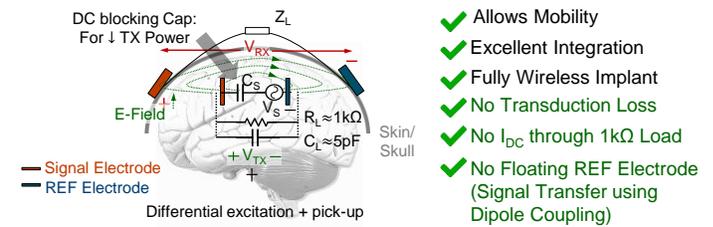
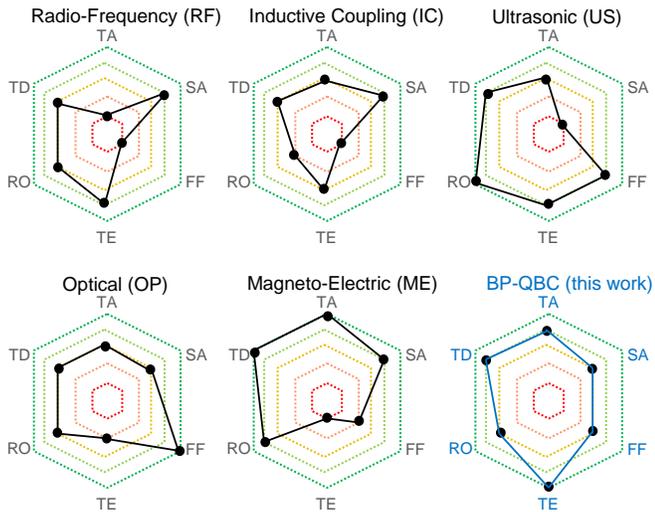
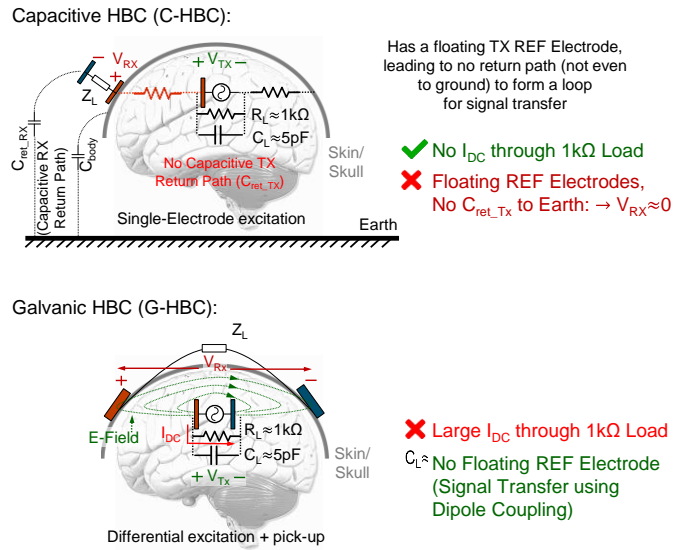
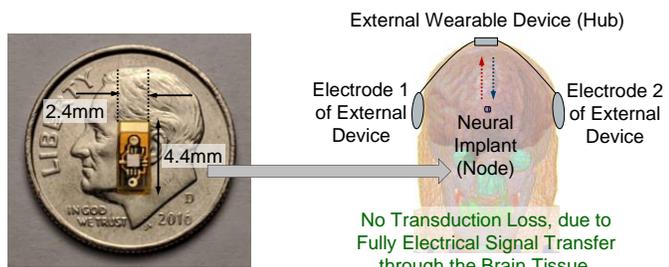
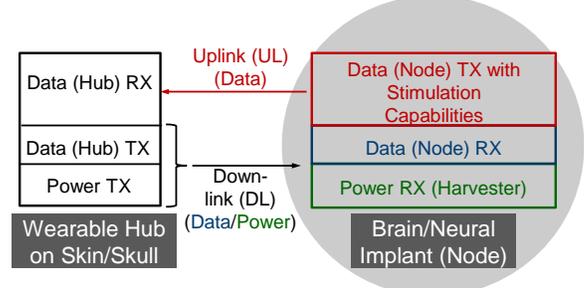

**Fig. 2 | Biphasic Electro-Quasistatic Brain Communication. a,** State-of-the-art Wired Brain Implants (an example is taken from Neuralink V2.0[5]). **b,** Wireless Signal Transfer mechanism of the proposed Bi-Phasic Quasistatic Brain Communication (BP-QBC) modality, along with its advantages. **c,** State-of-the-art of Wireless Brain Implants with untethered communication and powering, showing a performance comparison of Radio-Frequency (RF), Inductive Coupling (IC), Ultrasound (US), Optical (OP), Magneto-Electric (ME) and the proposed BP-QBC technique, demonstrating the high transduction efficiency with low tissue/skull absorption, large transmitter depth, and reasonable form-factor as well as robustness for BP-QBC. **d,** Comparison of the signal transfer mechanism with other Human Body Communication (HBC) modalities. Specifically, comparison with Capacitive HBC (C-HBC) and Galvanic HBC (G-HBC) is performed. **e,** Implementation of the 5.5 mm$^3$ BP-QBC System on a Chip (SoC) for wireless power and data transfer from a brain implant (node) to an external headphone-shaped hub placed on the skull/skin. **f,** System Architecture of the wearable hub and the neural implant (node). The Hub contains an uplink (UL) Data RX, a downlink (DL) Data TX and a DL Power TX. The node contains an UL Data TX, a DL Data RX and a DL Power RX for Communication and Powering. *Part of the human figures were created using the human model from* "NEVA EM *LLC*"[46] *and using the open-source software* 'MakeHuman'[49].

## BP-QBC: Analysis for Brain Implants

The system architecture for the wearable headphone-shaped hub and the BP-QBC-enabled brain implant is shown in Fig. 2f. The wearable hub contains an UL data RX, a DL data TX and a DL Power TX, while the brain implant consists of an UL data TX, a DL data RX, a bi-phasic stimulator and Energy harvesting modules. Communication from the node to the hub forms the UL, while communication from the hub to the node forms the DL. To properly characterize the BP-QBC signal transmission properties across different frequencies and channel lengths, an extremely small-sized (a few mm$^3$ or less) TX is required as an implant, which can sweep across kHz-GHz frequencies. Furthermore, the ground isolation requirements[18,31,34] in an HBC TX demands the TX to be self-sustained and not plugged in to any ground-connected power/signal source or measurement device. Due to such stringent volume, energy and ground-isolation requirements of a realistic brain implant, the core circuitry of the node in the said architecture is implemented in the form of a 0.3mm$^3$ custom-designed Integrated Circuit (IC), consuming only 1.15 µW, which can sweep across different frequencies, unlike commercially available signal sources. The details of the custom-designed IC can be found in our recent work[44]. Fig. 3a exhibits the die micrograph of our implemented IC, while Fig. 3b shows a conceptual diagram of the brain implant PCB implemented in this work. The Flexible Polyimide PCB consists of the 0.3mm$^3$ IC and a signal electrode on the front side, with a 3.8mm$^3$ storage capacitor and a reference electrode on the back side. Placing the electrodes on two different sides of the PCB helps in maximizing the distance between the UL TX electrodes, which is beneficial in terms of the channel transfer function (TF) as will be seen later in this section. The node is capable of sensing the neural signals, and communicating using BP-QBC as well as G-HBC. For the rest of this article, we shall focus mostly on the scientific mechanism of BP-QBC, and support it with theory and measurements.

### Signal Transmission Properties of the Brain Tissue

The electrical signal transfer properties in the human brain are shown in Fig. 3c, which exhibits the Relative Permittivity and Conductivity of Brain Tissue (both Grey Matter and White Matter) as a function of Frequency, showing that the tissue has a higher permittivity at low frequencies (kHz-MHz), which means that the signal transfer through the tissue mostly happens through creation of electric fields (dipole coupling where differential excitations are involved). The brain tissue has a high conductivity at higher frequencies (~1 GHz or more), where the signal transfer will happen mostly through electrical currents. However, at such frequencies, the electro-quasistatic approach does not hold true, and require considerations on signal reflection, constructive/destructive interference, and absorption of the EM fields in the tissue.

### Analytical Modeling from Dipole Coupling Theory

To develop an intuitive understanding of the signal transfer mechanism and the channel loss, we modeled the channel transfer function (TF) using the theory of dipole coupling in a 3-dimensional (3D) space. The channel TF as a function of the UL TX and RX geometries is shown in equation (4). A complete derivation is available in the supplementary materials.

$$\text{TF} = Q_{CD}\left[\frac{1}{r_{CA}-r_{e,C}} - \frac{1}{r_{CB}-r_{e,C}} - \frac{1}{r_{DA}-r_{e,D}} + \frac{1}{r_{DB}-r_{e,D}}\right] \bigg/ Q_{AB}\left[\frac{1}{r_{e,A}} - \frac{1}{r_{AB}-r_{e,A}} - \frac{1}{r_{AB}-r_{e,B}} + \frac{1}{r_{e,B}}\right] \quad (4)$$

where $A$, $B$, $C$ and $D$ are the center points of the node TX signal electrode, node TX reference electrode, hub RX signal electrode and hub RX reference electrode respectively (please note that the electrodes are assumed to be metal spheres for simplicity of analysis) as shown in Fig. 3d, and also later in supplementary Fig. 7; $r_{XY}$ represents the linear distance between points $X$ and $Y$ (for example, $r_{CA}$ is the linear distance between points $C$ and $A$); $r_{e,X}$ represents the radius of the electrode with a center point at $X$ (for example, $r_{e,A}$ is the radius of the TX signal electrode). Please note that *the channel TF is purely a function of the geometries, and is independent of frequency, as expected in the EQS regime*. equation (4) is further validated with Finite Element method (FEM)-based simulations in Ansys High Frequency Structure Simulator (HFSS), and the simulation results as a function of the channel length match closely with the analytical expression as shown in Fig. 3d. For a channel length of ~50-60 mm (realistic worst-case distance between a human brain implant and the hub electrodes), the loss is ~60dB from both analytical expressions and HFSS simulations. The channel TF for (1) various node (UL) TX electrode separation ($r_{AB}$) and (2) various radii of the node (UL) TX electrodes ($r_{e,A} = r_{e,B}$) are plotted in Fig. 4a-b, showing excellent conformity between the theory and simulation results. The RX electrode size does not have a significant effect in the signal transfer as the channel length is usually much higher than the electrode size. This was validated separately in simulation. Please note that for the analysis presented above, the implant is assumed to be equidistant from the hub electrodes, which results in a worst-case transfer function due to positional (as well as geometric) symmetry. For all other cases, the received signal will be higher, facilitating the signal transfer. However, for the current analysis, it is also assumed that the implant and the hub electrodes are perfectly aligned. This, however, results in the maximum received signal. To ensure such alignment of electrodes for all implants in the brain (even for the ones tilted at an angle), a skull cap with multiple electrodes can be used as the wearable hub for better signal quality at the RX in more realistic scenarios.

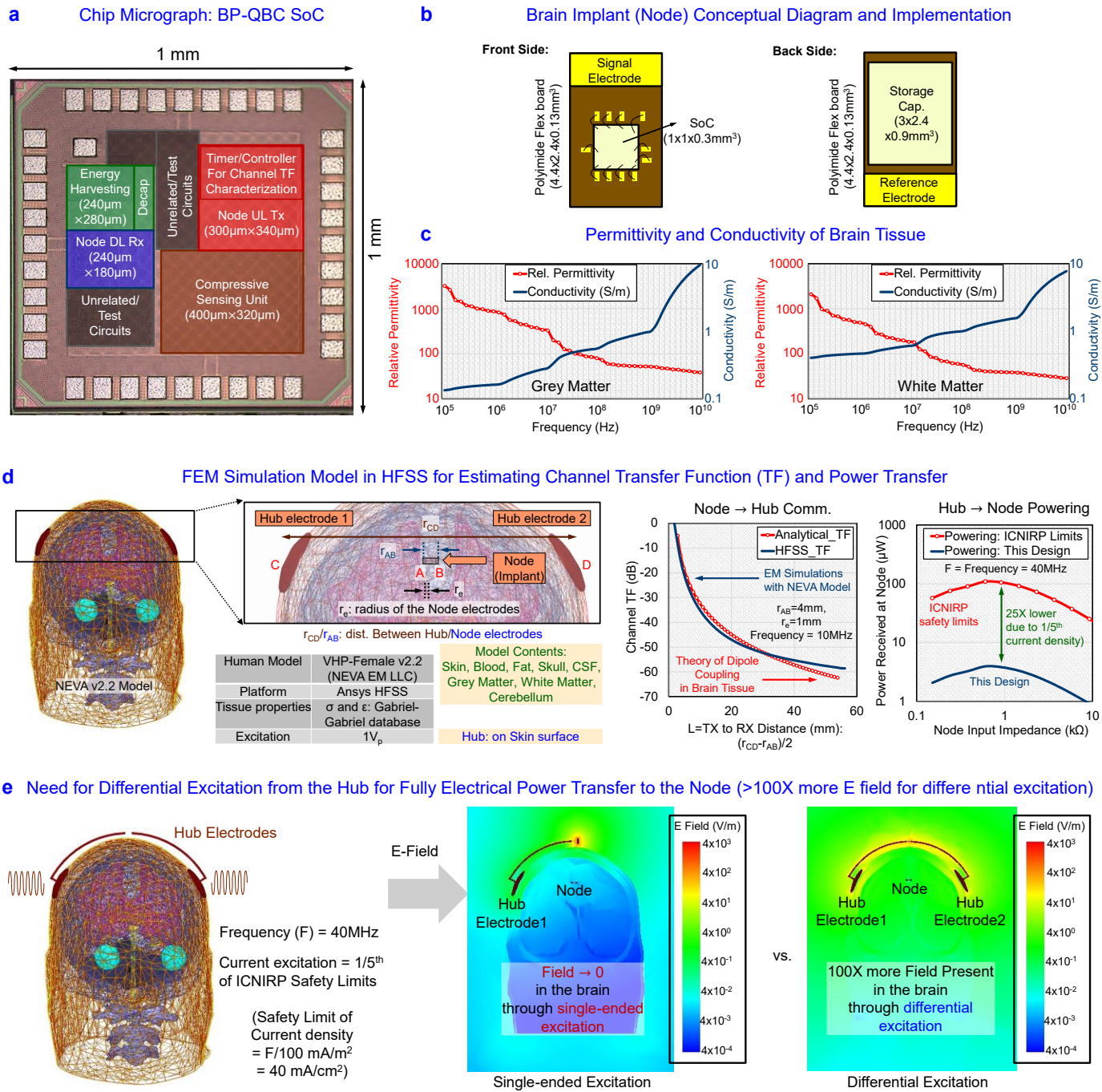

**Fig. 3 | Implementation and Modeling of Biphasic Electro-Quasistatic Brain Communication. a,** Die Micrograph of the BP-QBC SoC, showing Energy-Harvesting Modules, Node UL TX, Node DL RX and Hub UL RX. **b,** Conceptual Diagram of the SoC, placed on a flexible polyimide printed circuit board (PCB). **c,** Relative Permittivity and Conductivity of Brain Tissue (both Grey Matter and White Matter are shown) as a function of Frequency, showing that the tissue has a higher permittivity at low frequencies (kHz-MHz), which means that the signal transfer through the tissue mostly happens through creation of electric fields (dipole coupling where differential excitations are involved). The brain tissue has a high conductivity at higher frequencies (~1 GHz or more), where the signal transfer will happen mostly through electrical currents. However, at such frequencies, the electro-quasistatic approach does not hold true, and require considerations on signal refelection, and constructive/destructive interference. **d,** Finite-Element Method (FEM) based modeling using High Frequency Structure Simulator (HFSS), showing the channel transfer function (TF) for communication and powering for the device sizes of interest. The simulation results with a human model from NEVA EM LLC[45] shows excellent correspondence between simulation results and the theory of dipolse coupling for the Channel TF. When a 40MHz current excitation is used for powering, with current levels at $1/5^{th}$ of the ICNIRP safety limits[14-16] (i.e. 40mA/cm$^2$ from the hub), about 4μW of power is available at the implant. **e,** FEM simulation results showing creation of ~1000X stronger electric fields within the brain when powered from a hub with differential excitation as compared to a hub with single ended excitation for powering.

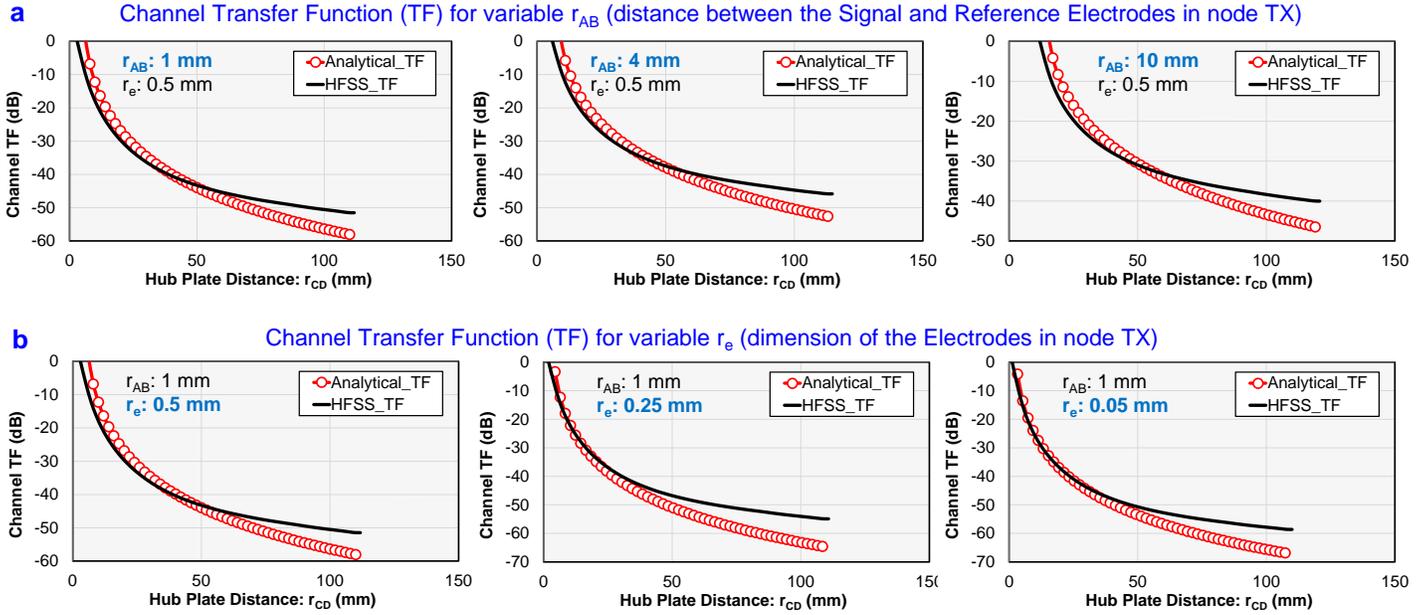

**Fig. 4 | Channel Transfer Function (TF) for the BP-QBC Implant using Analytical Expressions vs HFSS simulations with the human head model from NEVA EM LLC[45]. a,** Channel TF with variable distance between the distance of the transmitter electrodes ($r_{AB}$). **b,** Channel TF with variable dimension/radius of the transmitter electrodes ($r_{e,A}$).

**Powering the BP-QBC Implant**

The hub can also transfer power to the brain implant by differentially exciting the tissue using EQS signals. This is shown in Fig. 3e. According to the ICNIRP safety guidelines[14], a 40 MHz differential current excitation of 8 mA/cm$^2$ is used which is 1/5-th of the safety limits at 40 MHz (within EQS range). Such differential excitation results in > 100× higher electric fields within the brain tissue as compared to single-ended excitation (Fig. 3e), and this results in ~4 µW of available power at the implant sized ~5 mm$^3$, when the implant's input impedance is matched near the tissue impedance, as shown in Fig. 3d.

## Results

The main goal of performing the following experiments is to analyze the BP-QBC transfer function for in-vitro/in-vivo channels, as a function of the frequency and implant to hub distance. The simplified block diagram of the custom IC used for this purpose is shown in Fig. 5a, which has the capability of sweeping frequencies in the kHz-GHz range.

**In-Vitro Experiments: Channel TF as a Function of Frequency and Implant to Hub Distance**

Fig. 5b shows the measurement setup and methods for characterizing BP-QBC Channel TF as a function of (1) Frequency, and (2) Implant to Hub distance. The IC (1 mm × 1 mm × 0.3 mm) is placed on a Flexible PCB with dimensions 4.4 mm × 2.4 mm, and is submerged in Phosphate Buffered Saline (PBS) placed in a hemispherical plastic (polyethylene terephthalate - PET) bowl of ~60 mm radius (similar to the human skull). The entire setup is hung from the roof to minimize parasitic capacitive coupling to the earth's ground and nearby objects, thereby enhancing ground isolation. Two differential electrodes placed on the sides of the bowl work as the RX electrodes, which are connected to a TI BUF602 buffer configured as a 50 Ω driver. The TI buffer offers capacitive termination at the input of the RX, which is essential for establishing a wideband HBC channel as shown in earlier works[18,31,33,34] (on the other hand, a traditional 50 Ω termination results in a high-pass channel). The output of the buffer goes to a handheld Spectrum Analyzer from RF Explorer. The IC sweeps through **a** BP-QBC Implant (Node) Architecture for Uplink (UL) Communication and Stimulation

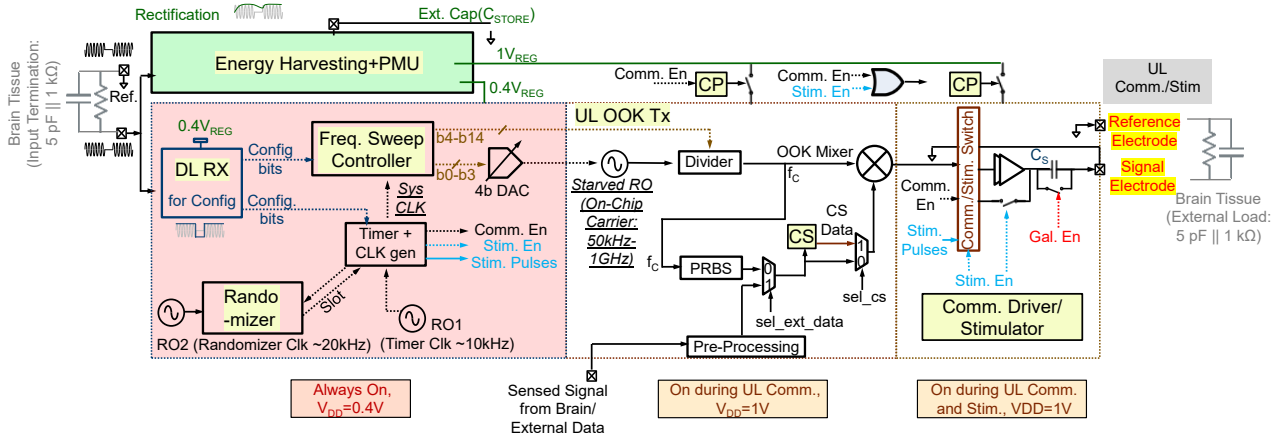

**b** In-Vitro Characterization Setup of BP-QBC Performance with Frequency and Channel Length

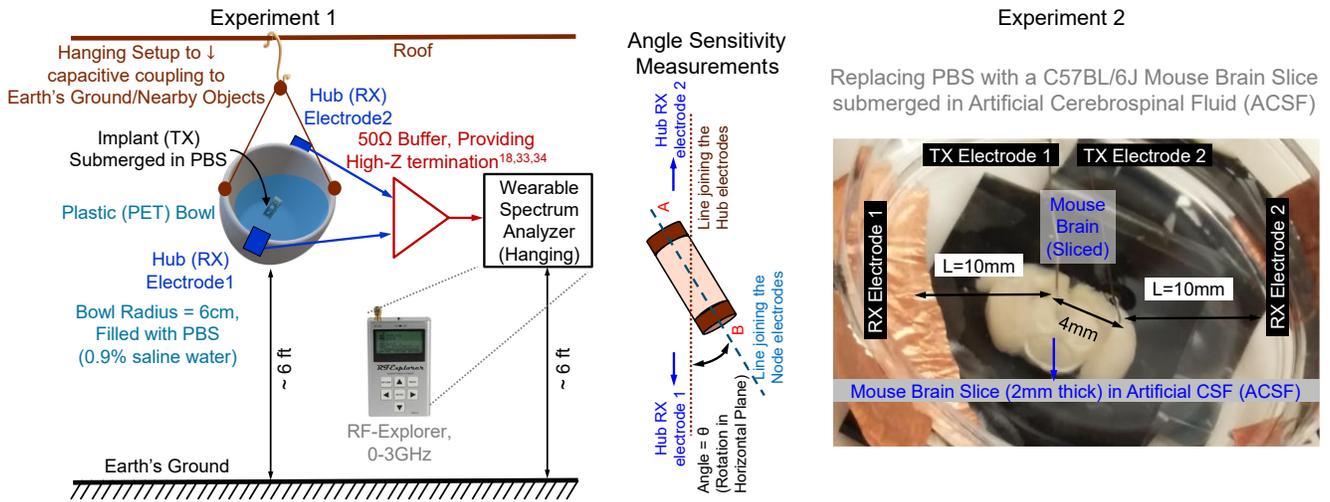

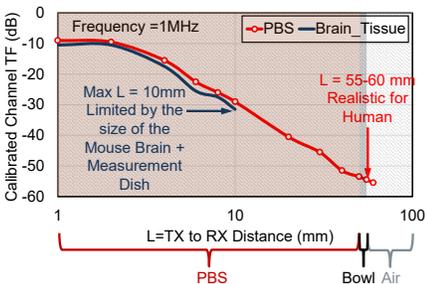

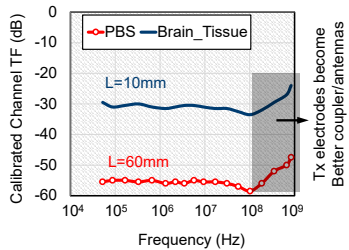

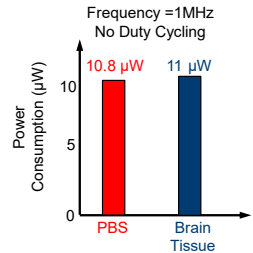

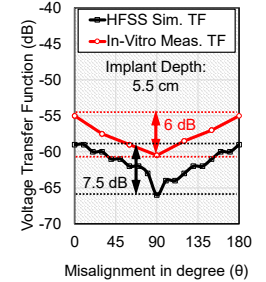

**Fig. 5 | BP-QBC Implant (Node) Architecture and Characterization. a,** The BP-QBC Implant Architecture for UL Communication and Stimulation, containing en energy-harvesting + Power Management Unit (PMU), a DL RX for configuring the Implant, a frequency sweep controller to sweep the output frequency for channel TF characterization (and to transmit at a particular freuqency), a randomizer to transmit the data for a randomized time slot specific to each implant, a compressive sensing module to compress the acquired data, and a configurable Comm./Stim Switch+driver, capable of G-HBC transmission, BP-QBC transmission and stimulation. **b,** Setup for characterizing BP-QBC performance with frequency and channel length, with the implant submerged in Phosphate Buffered Saline (PBS) is a hanging PET bowl to minimize capacitive coupling to earth's ground and nearby objects. By replacing the PBS with a mouse brain slice submerged in Artificial Cerebro-Spinal Fluid (ACSF), characterization on brain tissue is performed. **c,** Measurement results for channel TF vs. Channel length, showing reasonable correspondence between the results with PBS and brain tissue. For the mouse brain in ACSF, the maximum channel length was limited to only 10 mm (due to the size of the brain), while measurements up to 60mm were taken in PBS, showing a channel TF of about -60dB, which matches with our simulation results in HFSS. **d,** measured channel TF vs frequency, showing a flat response in both PBS and brain tissue, consistent with the electro-quasistatic approximation in the kHz-10's of MHz region. At higher frequencies, the electrodes start becoming better antennas, and electromagnetic (EM) effects can reduce the loss to some extent. **e,** Power Consumption in PBS and in the brain tissue at 1MHz, without any duty cycling, using the BP-QBC method. Using traditional G-HBC methods, the power consumption is ~41X higher at 1MHz frequency. **f,** Angle Sensitivity of the data link during uplink (UL) communication, showing ~6 dB worst case degradation in channel voltage TF in In-Vitro measurements, when the line joining the implant electrodes is at a 90° angle with the line joining the hub electrodes (please refer to Fig. 5b).

frequencies in the range of 40 kHz – 1 GHz to characterize BP-QBC signal transmission as a function of the Frequency. For characterizing the Channel TF as a function of the implant to hub distance, the RX electrodes are placed inside the bowl, and are moved at different distances from the implant. The experiments are repeated with brain slices from the C57BL/6J Mouse strain, adhering to Purdue University's overseeing Animal Care and Use Committee guidelines. Multiple brain slices, 500 μm – 2 mm thick, are placed in a measurement dish containing artificial CSF (ACSF) saturated with carbogen (95% $O_2$ + 5% $CO_2$). Two differential electrodes from the TX are placed on the surface of the brain slices.

Fig. 5c shows the channel TF as a function of the channel length (L = implant to hub distance), exhibiting ~30 dB loss for L = 10 mm and ~55 dB loss for L = 60 mm, at a frequency of 1 MHz. For the experiments with the mouse brain slice in ACSF, L is limited to ~10 mm due to the size of the brain slice. However, both the experiments with PBS and brain slice demonstrate similar channel TF as a function of the channel length, L. Also, there is no major discontinuity as the receiver moves from inside the bowl to outside, because in the EQS range of frequencies, the channel TF is primarily a function of geometries and transmitter to receiver distance, and not a strong function of the material. Fig. 5d shows the channel TF as a function of the frequency, for 2 scenarios – (1) with PBS (L = 60 mm), and (2) with mouse brain slice (L = 10 mm). The channel below 100 MHz range is almost flat-band, with a channel loss of ~55 dB for L = 60 mm and a channel loss of ~30 dB for L = 10 mm. The flat-band nature of the channel is consistent with the dipole coupling theory presented earlier in equation (4) for EQS communication. At higher frequencies ( > 100 MHz), the TX electrodes start working as better antennas, and the brain tissue/PBS start becoming more conductive, leading to a lower channel loss as seen in the right-hand side of Fig. 5d. However, at these frequencies, the signal transfer mechanism starts deviating from EQS, and becomes more EM-like, thereby leading to more radiation and tissue absorption. As shown in Fig. 5e, the power consumption of the IC without any duty cycling as ~11 μW at 1 MHz for both PBS and brain slice experiments. Finally, Fig. 5f shows the Angle Sensitivity of the data link during uplink (UL) communication, showing ~7.5 dB worst case degradation in channel voltage TF in simulation, and a ~6 dB worst case degradation in channel voltage TF in In-Vitro measurements, when the line joining the implant electrodes is at a 90° angle with the line joining the hub electrodes. For the In-Vitro measurements, once the implant is submerged in PBS, we only move the location of the receiver (hub) electrodes to change the angle with the implant, to create minimum disturbance to the submerged node.

**In-Vivo Experiments: Signal Acquisition and Power Consumption**

Fig. 6a shows the measurement setup and methods for characterizing BP-QBC Channel TF as a function of (1) Frequency, and (2) Implant to Hub distance. The IC (1 mm × 1 mm × 0.3 mm) is placed on a Flexible PCB of dimensions 4.4 mm × 2.4 mm, and is placed on the brain of a live C57BL/6J mouse, adhering to Purdue University's overseeing Animal Care and Use Committee guidelines. The mouse is anesthetized using a 2–3% isoflurane solution and is placed on a stereotaxic frame so that the head does not move during the experiment. Two differential electrodes placed on the sides of the skull work as the RX electrodes, which are connected to a TI BUF602 buffer configured as a 50 Ω driver, capacitively terminated at the input side. The output of the buffer goes to a handheld Oscilloscope from RF Explorer.

Fig. 6b shows the two modes that the IC (implant/node) can be configured – (1) G-HBC mode and (2) BP-QBC mode. In G-HBC mode, there is no DC blocking capacitor in the series path to prevent DC currents to go in the tissue, when the signal is not DC balanced. Since the IC generates a digital signal with on-off-keying (OOK), the output is not inherently DC balanced, and hence there will be a significant difference in power consumption in G-HBC and BP-QBC modes. This is shown in Fig. 6b. The power consumption for G-HBC flattens out for low frequencies (< 10 MHz) because of the ~1 kΩ or lower resistive load presented by the tissue, while BP-QBC power consumption continuously keeps on scaling with frequency. At a nominal quasistatic frequency of 1 MHz, BP-QBC consumes only ~11 μW power, as compared to the ~460 μW power consumption of G-HBC, thereby offering 41× lower power than G-HBC.

## Discussion

**BP-QBC UL Hub RX Requirements and Architecture**

A custom-made BP-QBC receiver (as part of the headphone-shaped hub), is utilized to test the BER of BP-QBC technique over data rates and input signal levels of the receiver as shown in Fig. 6c. This is implemented as an integrating receiver, which will be a part of the BP-QBC UL hub architecture. A Front-End (FE) integrating amplifier integrates the BP-QBC inputs, which is followed by a passive envelope detector (ED) for OOK demodulation, and an integrator follwed by a sampler that uses the clock from an oversampled clock and data recovery (CDR) unit. The need for an integrating receiver arises to *compensate for the high-pass-filtering/differentiating effect* introduced by the series capacitance $C_S$ in the BP-QBC transmitter (the implant), in conjunction with the tissue load resistance, $R_L$ (Fig 6b and Fig. 6d). The TSMC 65 nm ASIC implementation is shown in Fig. 6e, while the bit-error-rate (BER) vs. input voltage to the Hub Receiver for data rate (DR) = 10 Mbps is demonstrated in Fig. 6f. This experiment is performed with a carrier frequency of 100 MHz, which is near the boundary of EQS and EM range of frequencies, as shown by the channel TF of Fig. 5d. For a target BER of $10^{-3}$ at 10 Mbps DR, the required input voltage level at the receiver for the current experiment is about -52 dBV (~2.5 mV), which means that even for a channel loss of ~50 dB, the transmit voltage level of 1V should be sufficient for achieving BER ~ $10^{-3}$ at DR = 10 Mbps. However, for higher channel losses (larger implant depth), we shall need to use lower DR to achieve similar BER. The data compression unit (CS) implemented in the BP-QBC node TX helps significantly for such high loss scenarios. With a 33.33× data compression, 300 kbps DR through the channel can effectively represent the same 10 Mbps communication. The sensitivity of the receiver can then be improved from -52 dBV (no compression, 10 Mbps, 100 MHz carrier) to -68 dBV (33.33× compression, 300 kbps, 30 MHz EQS carrier) for a target BER of $10^{-3}$, as shown in Fig. 6f.

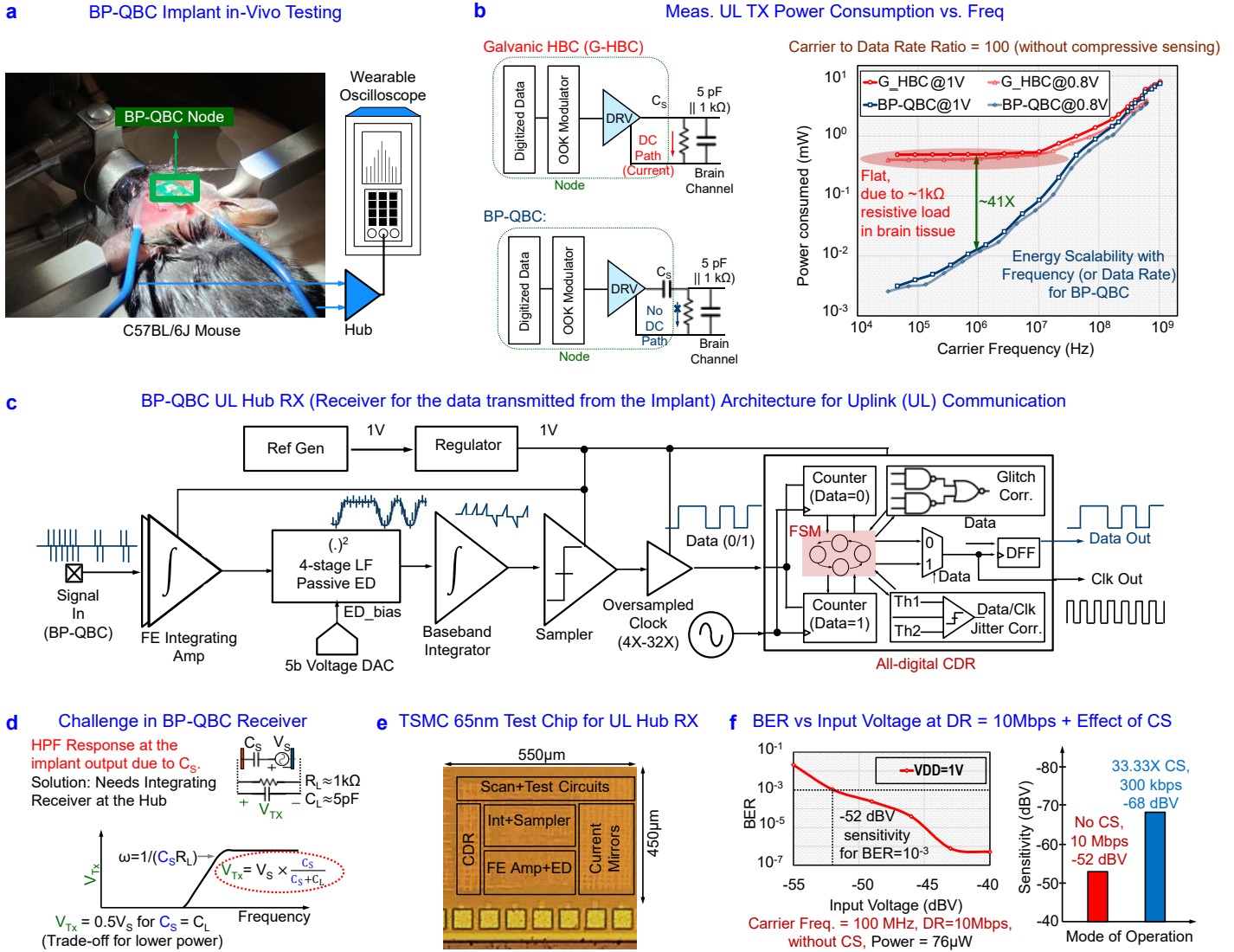

**Fig. 6 | Measurement Results from the BP-QBC SoC. a,** The C57BL/6J Mouse, with the Node placed on its brain. **b,** Comparison of UL Power Consumption in G-HBC modality vs BP-QBC. **c,** BP-QBC UL Hub RX Architecture, showing the integrating receiver. The Front-End (FE) amplifier integrates the BP-QBC inputs, which is followed by a passive envelope detector (ED) for OOK demodulation, and an integrator with a sampler that uses the clock from an oversampled clock and data recovery (CDR) circuit. **d,** The need for an integrating receiver arises to compensate for the high-pass-filtering/differentiating effect introduced by the series capacitance $C_S$ in the BP-QBC TX (the implant), in conjunction with the tissue load resistance, $R_L$. **e,** The TSMC 65nm test chip implementation for the Hub RX[50]. **f,** BER vs. input voltage to the Hub Receiver for data rate (DR) = 10Mbps without compression.

**Table 1 | Comparison of Bi-Phasic Quasistatic Brain Communication (BP-QBC) with state-of-the-art techniques for wireless communication in a Brain Implant, including RF, inductive, Optical, Ultrasonic, Magneto-Electric and Traditional HBC Methods.**

| | RF Comm.[6,7,13] | Inductive Comm.[11,12] | Optical Comm.[8] | Ultrasonic Comm.[9] | Magneto-Electric Comm.[10] | C-HBC[36,38] | G-HBC (Ref. Design) | **THIS WORK** BP-QBC (Focus of This Work) |
|---|---|---|---|---|---|---|---|---|
| **External Components** | Coil + Capacitors | Coil + Capacitors | LED + PV Diode + Electrodes | Piezo Device + Electrodes | ME Film + Capacitor + Electrodes | Battery + Electrodes | Capacitor + Electrodes | **Capacitor + Electrodes** |
| **Sub-Cranial Repeater/Relay** | Yes | No | Yes | Yes | No | No | No | **No** |
| **Node To Hub Distance** | ~10 mm | ~10 mm | ~20 mm | ~50 mm | ~30 mm | ~10 mm | ~60 mm | **~60 mm** |
| **End-To-End Channel Loss (60 mm epth)** | High ($> 80$ dB) | High ($> 80$ dB) | High ($> 80$ dB) | High ($> 80$ dB) | Moderate ($\approx 80$ dB) | High ($> 80$ dB) | Low ($\approx 60$ dB) | **Low ($\approx 60$ dB)** |
| **Power Consumption** | High ($> 500$ μW) | High ($> 100$ μW) | Very Low ($< 1$ μW) | Low ($< 50$ μW) | High ($> 500$ μW) | Very Low* ($< 5$ μW) | Low* ($< 10$ μW) | **Very Low* ($< 1$ μW @1Mbps)** |
| **Max. Data Rate (bps)** | Moderate (~1M) | Low (~100k) | Very Low (~100) | Very Low (~100) | Moderate (~1M) | High (>10M) | High (>10M) | **High (>10M)** |
| **Energy Eff. (pJ/b)** | Inefficient ($> 500$) | Inefficient ($> 1000$) | Inefficient ($> 1000$) | Inefficient ($> 10000$) | Inefficient ($> 500$) | Efficient (~50) | Inefficient ($> 500$) | **Efficient (~50)** |

* Estimated with 1% duty cycling at 10 MHz carrier frequency and 1 Mbps data rate, without any data compression

**Link Efficiency Estimation for BP-QBC**

Table 1 compares Bi-Phasic Quasistatic Brain Communication (BP-QBC) with state-of-the-art techniques for wireless communication in a Brain Implant, including RF, inductive, Optical, Ultrasonic, Magneto-Electric and Traditional Galvanic HBC Methods. Unlike RF, the absorption of EQS signals in the brain tissue is lower. Additionally, unlike optical or ultrasonic techniques, EQS is not affected much by the presence of the skull (which necessitates sub-cranial repeaters/relay units for optical and ultrasonic communication). This, along with the fact that BP-QBC does not require any transduction of energy (unlike optical, ultrasonic or magneto-electric techniques), allow us to have only ~60dB end-to-end channel loss as found from our analysis, simulations and measurement results. Even though the power consumption for optical and ultrasonic methods can be made extremely low, the data rates are usually limited by the sensitivity of the RX which depend on back-scattered signals. Conversely, HBC techniques (such as C-HBC, G-HBC and BP-QBC) allow high data rates with low power with proper duty cycling. As explained earlier, BP-QBC has a system TF similar to Galvanic HBC, with a power consumption similar to Capacitive HBC, thereby making it one of the most promising techniques of communication with a brain implant. It could, however, be argued that optical and ultrasonic techniques offer better spatial resolution due to the smaller wavelengths, and leads to miniaturized implants. BP-QBC, however, offers a much better end-to-end system loss, allowing room for further miniaturization as compared to the implementation presented in this paper. The analysis and comparison of the end-to-end system loss for different techniques is presented in the supplementary notes.

**Privacy Space Comparison for Security: BP-QBC vs Traditional Methods**

In addition to low power consumption, high data rates, and no transduction loss, BP-QBC also promises to exhibit significant security benefits for an implant. Since the EQS signals do not radiate significantly outside the human body, all EQS HBC modalities are inherently more secure than RF techniques where the signals leak outside the body and can be picked up by an attacker at a distance of a few cm – a few m. Our previous works[42,45] have explored the security properties of G-HBC and C-HBC in detail, and compared them with RF to show that EQS HBC techniques exhibit a private space which is ~30× better[45] than traditional radiative RF, thereby making it harder for a malicious attacker to snoop in. A detailed analysis of the security properties of BP-QBC, and comparison with the security of OP, US and ME techniques is out of scope of the current article and will be analyzed in a future work.

## Conclusions

This work, for the first time, analyzes fully-electrical quasistatic signaling for Wireless Communication from a Brain Implant, and demonstrates the *first BP-QBC link* with *simultaneous powering and communication*. EQS (wherein the signal wavelengths are much larger than the channel length through the body) is preferred over EM operation because of the lower power consumption at the TX (due to lower frequency of operation and the ability to have high-impedance terminations), lower absorption in the brain tissue and lower leakage/radiation in the surrounding media, which makes the data transmission more secure, as an attacker cannot snoop into the data through a wireless EM channel. Due to *no field transduction*, the end-to-end channel loss in BP-QBC is only ~60 dB at a distance of ~55 mm with a 5 mm$^3$ implant, which is > 20 dB better than competing techniques (optical/ultrasound/ magneto-electric), and *allows room for further miniaturization of the node*. Understanding that the EQS signal transfer through the brain channel occurs through AC electric fields, while the primary source of power consumption is due to galvanic DC currents arising from the finite conductivity of brain tissues, we utilized a DC-blocking capacitor to block the DC current paths through the brain tissue without significantly affecting the bi-phasic AC communication at EQS frequencies. Due to this, the power consumption in BP-QBC is measured to be ~41× lower than traditional G-HBC at a nominal electro-quasistatic frequency of 1 MHz. Furthermore, unlike optical and ultrasonic techniques, BP-QBC *does not require sub-cranial interrogators/repeaters* as the EQS signals can penetrate through the skull and has enough strength due to the low loss channel, which makes it an extremely promising technique for *high-speed, low-loss data transfer through the brain tissue with harvested energy limits*.

## Methods

This section provides the details related to our simulation and experimental methods, to facilitate reproduction of the results by another independent researcher.

### Setup for FEM Simulation

**Simulator and Models**: All the EQS simulations have been performed in Ansys High Frequency Structure Simulator (HFSS), which is a Finite Element Methods (FEM) based solver for Maxwell's Equations. A detailed human head model consisting of realistic tissues (white matter + gray matter + CSF + Skull + Blood + Skin) is used to validate the theoretical channel transfer function, which is taken from NEVA Electromagnetics[46]. Dielectric properties of the brain tissues have been obtained from the works of Gabriel *et al.*[21,22]

**Implant Model**: A simple cylindrical model made of rubber, along with two spherical copper electrodes, are used as the implant, as shown in supplementary Fig. 7. The nominal radius of this rubber cylinder is 0.5 mm, while the nominal height is 4 mm. two spherical copper electrodes of radius 0.5 mm are placed on the two sides of this model to represent the node TX electrodes. The rubber cylinder is curved at the sides to support the two electrodes and cover them in a hemispherical manner. This implant model is floated 6 cm within the human head/brain model, while the head model itself is floated 1.7 m above a plane with Perfect E-Boundary in HFSS, which replicates an infinite ground plane similar to the earth's ground. The Excitation for the simulation is provided through differential/galvanic coupling, as described in the next sub-section.

**Excitation:** A differential/galvanic coupling model is used to provide excitation to the brain tissue surrounding the implant. The coupler consists of two copper spheres with a radius of 0.5 mm. The separation between the two spheres (which is filled with the rubber cylinder, curved near the electrodes) can be varied, as well as the radius of the spheres and the cylinder. A voltage source excitation is placed between the two spheres. In HFSS, this imparts an alternating potential difference of amplitude 1 V between the two electrodes, replicating an ideal AC voltage source. This is unlike the traditional lumped port excitation method in HFSS, which is suitable for 50 Ω matched excitations, but may result in unexpected reflections when coupled with a non-standard termination model.

**Measurement of Voltage at the Hub Receiver:** The receiving node structure uses parallel discs of similar dimensions, placed on two sides of the head model, as shown in Fig. 3d. A lumped RLC boundary is placed between the electrode and the ground plate at the receiver, which is set to 10 pF for capacitive high impedance termination, modeling the capacitance between the electrode to local ground reference. The potential difference between the discs is calculated by performing a line integration of the electric field along the straight line between the receiver electrode and ground plates.

**Measurement of Power at the Node Receiver:** For DL powering purposes, the hub becomes the power TX and the implant becomes the power RX. The input impedance at the implant needs to be matched with the tissue impedance for maximum power transfer. A 40 MHz differential voltage source excitation is placed between the two hub electrodes, imparting an alternating potential difference of amplitude 1 V between the two electrodes, replicating an ideal AC voltage source. A lumped RLC boundary is placed between the electrodes of the implant, which is varied from R = 100 Ω to R = 10 kΩ. The peak potential difference (V) between the plates is calculated by integrating the electric field along a straight line between the electrode and ground plates, and the power received is calculated by the formula P = $V^2/2R$. For an AC amplitude of 1 V between the two electrodes at the hub, the effective rms current density at the DL Power TX electrodes is calculated to be $\frac{1}{\sqrt{2}}\frac{V}{R_{Tissue}} \times \frac{1}{\pi r_{e,C}^2}$ where $R_{Tissue}$ is the effective tissue impedance seen by the Power transmitter (~1 kΩ), and $r_{e,C}$ is the radius of the electrodes in the power transmitter/hub (~1cm). This results in an effective current density of 0.225 mA/cm$^2$, while the power transmitted becomes $\frac{1}{2}\frac{V^2}{R_{Tissue}} \approx 500$ μW. Using HFSS, FEM simulations are performed for worst case channel power TF, which is found to be about -45 dB for a maximally misaligned node at a depth of about 6cm, at a 90$^O$ angle with the line joining the hub electrodes (Please refer to the supplementary notes for more analysis on the misalignment). The power delivered to the 1 kΩ input impedance at the implant will be only 7.9 nW, taking into account the power division between the tissue and the implant's input impedance. However, according to the ICNIRP safety guidelines[14] (Please refer to Table 4, page 509 in the guidelines[14]), the allowable current density for human head in occupational scenarios is f/100 mA/m$^2$ for frequencies up to 10 MHz, where f denotes the frequency. This results in a maximum allowable current density of f/1M mA/cm$^2$, which is 10 mA/cm$^2$ at 10 MHz. By linearly scaling this to 40 MHz (which is also the human-body powering frequency used in previous works[35,37]), the allowable current density can be calculated to be 40 mA/cm$^2$. If we use 1/5$^{th}$ of this current density magnitude (8 mA/cm$^2$, which is also the limit for general public exposure), the power available at the 1 kΩ input impedance at the implant will be $7.9 \times \left(\frac{8}{0.225}\right)^2$ nW ≈ 10 μW, while the transmitted power is $500 \times \left(\frac{8}{0.225}\right)^2$ μW ≈ 632 mW. With an assumption that only 40% of the equivalent energy can be stored at the node, we estimate that only ≈ 4 μW of power can be harvested at the brain implant, which is considered as the target power budget of the implant. *The power transfer efficiency from the hub to the input impedance at the implant is thus ≈ 10 μW / 632 mW ≈ 0.0015% in the worst case.* However, when the node is aligned with the hub electrodes, the channel power TF is about -33 dB at an implant depth of 6 cm, which will lead to a *best-case power transfer efficiency of ≈ 0.025%, and in that case, the transmit power from the hub needs to be about 40 mW for 10μW available power at the node.*

**Setup for Characterization Experiments (In-Vitro/In-Vivo)**

**Node Design (Data Transmitter and Power Receiver):** The specifications of the implant/node for characterization experiments demanded a signal generator of small form factor (~5 mm$^3$) with proper ground isolation, which runs on a self-sustained energy source (battery/storage capacitor), and sweeps through a few kHz-1 GHz. For this purpose, an integrated circuit (IC) was built on a 1 mm × 1 mm × 0.3 mm System on a Chip (SoC) fabricated in TSMC 65 nm complementary metal-oxide-semiconductor (CMOS) technology, which serves as an ultra-low power implantable signal generator. The implemented BP-QBC SoC (Fig. 5a) is equipped with (1) a 52 pJ/b energy-frequency scalable UL TX with on-chip clock, on-off keying (OOK)-based modulation, and variable duty-cycling, along with compressive sensing (CS) for data volume reduction, and collision avoidance while sending data from multiple implants; (2) an always-on 31 nW DL RX to receive system configuration bits and control signals from the wearable hub; (3) a bi-phasic stimulator with 89.2% current efficiency ($I_{Stim}/I_{DC}$ = ratio of current supplied for stimulation with the consumed DC current); and (4) an energy-harvester utilizing a 30-stage RF-rectifier[47] (RR) that can generate a 1 V supply for the SoC with only ~70 mV$_{Peak}$ DL received input. Two different supply domains are implemented: a 0.4 V domain for low-leakage/low-power always-on timer/controller modules, and a separate 1 V domain for duty-cycled data communication and stimulation. The power management unit in the SoC consists of a 13 nW reference voltage generator (for both 0.4 V and 1 V supplies) and two 24 nW low-dropout regulators (LDO) that generate the $V_{DD}$ for the SoC, utilizing the energy harvested from the 30-stage RF-Rectifier. In the current implementation, the SoC utilizes 1:100 duty cycling with a 100 ms long transmit phase, and a 100 ms long stimulation phase within a total time of 10s by default. Using the DL control signals, additional modes with 1:1000 and 1:10 duty cycling can also be configured, or duty cycling can be turned off. Fig. 5a presents a simplified diagram of the building blocks of the SoC. The external energy-storage capacitor ($C_{STORE}$) at the output of the RF-Rectifier is carefully optimized during design time for a maximum data rate (DR) of 10 Mbps with a charging time ≪ 100 s and a voltage droop ≪ 100 mV during each of the transmit and stimulation phases. A 17 nW charge pump (CP) generates an output voltage ($V_{PUMP} \approx 1.8V$) which is much higher than the $V_{DD}$ (1V), and is utilized to bias specific power gates on the supply of the duty-cycled modules in deep-subthreshold during off-state to reduce their leakage. Using such bootstrapping techniques, the leakage at the power gates is reduced from ~0.51 μW to ~1 nW (> 500× reduction). The average power consumption in the SoC is only 1.15 μW (including leakage) with 1% duty cycling, out of which 0.52 μW is consumed in the BP-QBC driver, at 3 MHz carrier frequency (1 Mbps effective data rate with 33.33× compression). A wake-up controller based compressive sensing (CS) front-end allows compression of any acquired data in the neural sensor to reduce the overall TX data rates. The CS module consists of an on-chip 2-stage discrete wavelet transform (DWT)-based optional sparsifier and a dual varying-seed-PRBS-based sensing-matrix generator, an allows a variable compression factor (CF) in between 5× to 33.33×. A ring-oscillator based physical unclonable function (RO-PUF) designed as a 9-bit pseudo random bit sequence (PRBS) generator is utilized to specify randomized time slots for transmission and stimulation in different nodes within the brain. This enables an inherent collision avoidance scheme without any complex medium access control (MAC) implementation. The DL RX in the implant SoC consists of a 10.1 nW Front-End (FE) amplifier, a 3.2 nW 4-stage passive envelope detector for demodulation of configuration bits, and a 16.2 nW fully digital oversampled clock and data recovery (CDR) circuit for data decoding. Further details on the implementation can be found in our recent works[44,48].

**Need for Compressive Sensing (CS):** Since neural signals can range from very low frequency (few mHz – 10's of Hz) local field potentials to higher frequency (10's of Hz – a few kHz) action potentials, the data rates for a single channel can reach a few 100's of kbps. As an example, a signal acquisition module with 10 kHz bandwidth, 5× oversampling, and 16 bits per sample results in a data rate of 10k×5×16 = 800 kbps. For multi-channel signal acquisition, this requirement increases further. If a carrier to data rate ratio of 100 is utilized for data transfer using OOK modulation, the power consumption becomes ~770 µW for BP-QBC and ~1120 µW for G-HBC (please refer to Fig. 6b), requiring aggressive duty cycling for communicating within the ~4 µW of available power limit at the implant (please refer to Fig. 3d-e). On the other hand, a compressive sensing (CS) front-end can reduce the overall energy consumption per bit by ~16×, while reducing the data rate (and hence the required carrier frequency) by 33× as shown in our earlier works[44,48]. This results in an effective carrier to data rate ratio of only ~3, reducing the power consumption of BP-QBC to only ~26 µW for 800 kbps (~2.4 MHz) transmission. With a 10 MHz carrier, the energy efficiency of the BP-QBC system is 835 pJ/b without CS (with a carrier to data rate ratio of 100), which reduces to only 52 pJ/b with CS. If a carrier to data rate ratio of ~10 is allowed for proper detection of data at the RX, the energy efficiency with CS can reduce to < 10 pJ/b. At 1 Mbps data rate with compressive sensing (CF = 33.33×), the power consumption in the BP-QBC driver is only 0.52 µW with 1% duty cycling, which is within the range of the available power (~4 µW) at the implant using power transmission through the brain tissue.

**Need for Collision Avoidance:** As mentioned earlier, another important feature in the SoC is an inherent collision avoidance scheme implemented to avoid/reduce the chances of multiple nodes transmitting data at the same time (or stimulating at the same time) within the brain. Due to the small amount of available power and the high bandwidth requirement of neural signals, multiple nodes cannot operate simultaneously with frequency division multiplexing (FDM), and hence time-division multiplexing (TDM) needs to be used. To ensure that TDM can be implemented (1) without any medium access control (MAC) layer protocol, and (2) without any synchronization among multiple nodes placed within the brain, a physical unclonable function (PUF) based communication and stimulation slot selector is utilized in conjunction with duty cycling. 1% duty cycling theoretically allows 100 implants to operate simultaneously, while a 9-bit ring-oscillator based PUF ensures that there needs to be at least 27 nodes operating simultaneously so that at least two of the nodes transmit at the same time slot with 50% (or more) probability, which can be proven from theory.

**Setup for In-Vitro Experiments:** Fig. 5b demonstrates the measurement setup and methods for characterizing the in-vitro BP-QBC Channel TF as a function of Frequency as well as Implant to Hub distance. The 1 mm × 1 mm × 0.3 mm SoC is housed on a Flexible Polyimide PCB of dimensions 4.4 mm × 2.4 mm, and is subsequently submerged in Phosphate Buffered Saline (PBS) water placed in a hemispherical plastic (polyethylene terephthalate - PET) bowl of 60 mm radius, which has similar dimensions as that of the human skull. The entire setup is suspended from the roof to minimize any parasitic capacitive coupling to the earth's ground and nearby objects, thereby improving ground isolation during the measurements. Two differential metal electrodes attached to the sides of the bowl work as the RX electrodes, which are connected to a TI BUF602 buffer configured as a 50 Ω driver for measurement instruments. The TI buffer offers ~2 pF capacitive termination at the input of the RX, which helps in establishing a wideband HBC channel as shown in earlier works[18,31,33,34]. On the other hand, a traditional 50 Ω termination would have resulted in a high-pass channel. The output of the buffer is terminated with 50 Ω, and goes to a handheld Spectrum Analyzer from RF Explorer. The IC sweeps through different configurable frequencies in the range of 40 kHz - 1 GHz to characterize the BP-QBC signal transmission as a function of the Frequency. For characterizing the Channel TF as a function of the implant to hub distance, the RX electrodes are placed inside the bowl, and are subsequently moved at different distances from the implant. The experiments are repeated with brain slices from a C57BL/6J Mouse strain, adhering to the guidelines of the overseeing Institutional Animal Care and Use Committee (IACUC) at Purdue University. During this experiment, 500 µm – 2 mm thick slices are placed in a measurement dish containing artificial CSF (ACSF) saturated with carbogen (95% $O_2$ + 5% $CO_2$). Two differential electrodes from the TX are placed on the surface of the brain slices for excitation. The RX electrodes are placed at different distances from the brain slices for this experiment. A total of 7 instances of the IC were tested, and the worst-case results were reported. During the in-vitro experiments, the implant is powered from a pre-charged 3 mm × 2.4 mm × 0.9 mm CP3225A supercapacitor which is housed on the back-side of the Flexible Polyimide PCB.

**Setup for In-Vivo Experiments:** Fig. 6a exhibits the measurement setup and methods for characterizing the BP-QBC in-vivo Channel TF as a function of Frequency as well as the Implant to Hub distance. The 1 mm × 1 mm × 0.3 mm SoC is housed on a Flexible polyimide PCB of dimensions 4.4 mm × 2.4 mm, and is placed on the brain of a live C57BL/6J mouse. All the research protocols were approved and monitored by the Purdue University Institutional Animal Care and Use Committee (IACUC), and all research was performed in accordance with relevant NIH guidelines and regulations. The mouse was anaesthetized with 2–3% isoflurane throughout the surgery. After shaving the hair, the animal was fixed on a stereotaxic frame, so that the head does not move during the experiment, and the head skin was sterilized. Up to 2.5 cm sagittal incision was made in the skin over the skull, and a bilateral craniotomy was performed using a precision surgery dental drill. After the craniotomy, the skull in the midline was thinned down to improve contact with the BP-QBC implant. The insertion of the BP-QBC electrodes were manually done. Two differential electrodes placed on the sides of the skull work as the RX electrodes, and were fixed with super glue to keep them in place. The RX electrodes were connected to a TI BUF602 buffer configured as a 50 Ω driver, with a ~2 pF capacitive termination at the input side. The output of the buffer goes to a handheld Oscilloscope from RF Explorer with 50 Ω termination. During the in-vivo experiments, the implant is powered from a pre-charged 3 mm × 2.4 mm × 0.9 mm CP3225A supercapacitor which is housed on the back-side of the Flexible Polyimide PCB.

## Data Availability

The data that support the plots within this paper and other findings of this study are available from the corresponding author upon reasonable request.

## Code Availability

Custom codes used to process the data are available from the corresponding author upon reasonable request.

## Acknowledgements

This work was supported by the Air Force Office of Scientific Research YIP Award (FA9550-17-1-0450), the National Science Foundation CAREER Award (grant#: 1944602) and the National Science Foundation CRII Award (CNS 1657455). The authors would like to thank Dr. Debayan Das, PhD, Purdue University and Mr. Nirmoy Modak, Current PhD student, Purdue University for their co-operation and support during development of the integrated circuits for the neural node.

## Author Contributions

B.C. and S.S. conceived the idea, B.C., S.S., and M.N. contributed to designing the experiments for BP-QBC. B.C., M.N., G.K.K. and S.S. conducted the theoretical analysis, numerical simulations, node design and performed the characterization experiments. B.C., S.X., K.J. and G.K.K. performed the in-vitro and in-vivo animal experiments. B.C., S.S., M.N. and K.J. analyzed the experimental data. B.C., M.N. and S.S. wrote the paper. All the authors contributed to reviewing and revising the manuscript.

## Competing Interests

The authors declare that they have no competing financial and/or non-financial interests.

## Additional Information

**Supplementary information** is available for this paper at https://doi.org/10.48550/arXiv.2205.08540.
**Correspondence and requests for materials** should be addressed to B.C.

# Supplementary Information

## Notes 1: Derivation of the Analytical Channel Transfer Function (TF) Expression

This section provides a derivation of the expression as presented in equation (4), for the analytical channel transfer function.

Fig. 7 shows a simplified version of the BP-QBC implant model and electrodes for both the UL TX and UL RX. For simplicity of the derivation, we shall assume spherical metal electrodes at both the TX and the RX. As explained earlier in the description of equation (4), points A and B are the center points of the TX electrodes, while points C and D are the center points of the RX electrodes. $r_{XY}$ represents the linear distance between points $X$ and $Y$ (for example, $r_{AB}$ is the linear distance between points $A$ and $B$); $r_{e,X}$ represents the radius of the electrode with a center point at $X$ (for example, $r_{e,A}$ is the radius of the TX electrode 1 as shown in Fig. 7).

Please note that in the analysis presented next, the implant is assumed to be equidistant from the hub electrodes, which results in a worst-case transfer function due to positional (as well as geometrical) symmetry. For all other cases, the received signal will be higher, facilitating the signal transfer. However, for the current analysis, it is also assumed that the implant and the hub electrodes are aligned, as shown in Fig. 7. This, however, results in the maximum received signal. To ensure such alignment of electrodes for all implants in the brain (even for the ones tilted at an angle), a skull cap with multiple electrodes can be used as the wearable hub for better received signal quality in more realistic scenarios.

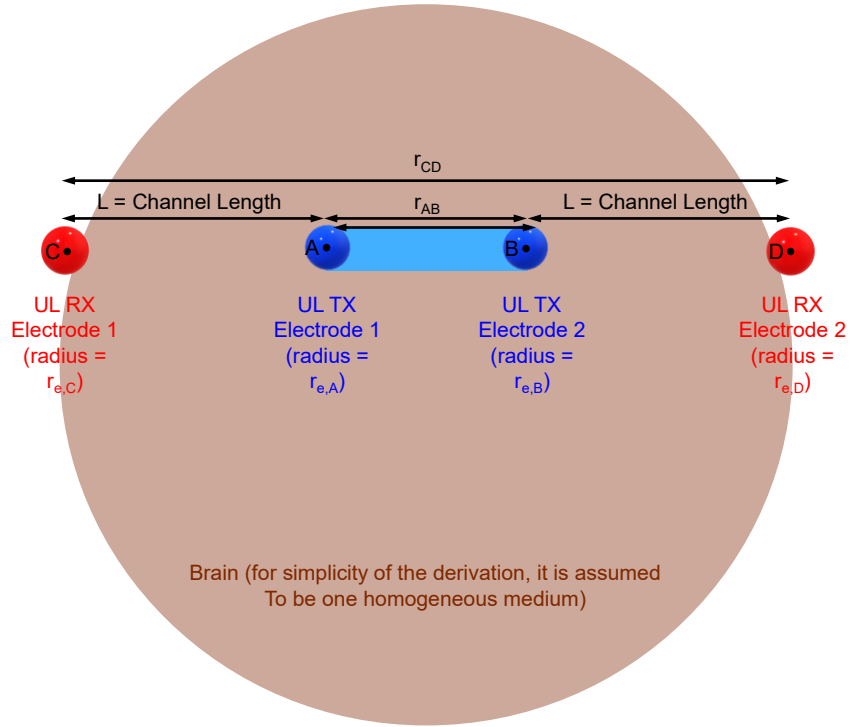

**Fig. 7 | Implant Dimensions and Positioning.** The UL TX (implant) as well as the UL RX (wearable hub) are assumed to have spherical electrodes for simplicity. The brain model is also a sphere, and assumed to be homogeneous for simplicity.

We start the analysis by calculating the voltage on a spherical resistive shell – first with a point charge placed at the center of the conductive shell, and then by considering the source to be a spherical metal electrode with radius $r_e$ instead of the point charge. This analysis helps finding the voltage at any point within the brain, due to a monopole (single) electrode.

As shown in Fig. 8a, for a point charge $+q$, the differential resistance $dR$ at a distance of $r$ can be calculated as equation (5):

$$dR = \rho \frac{dr}{A} \quad (5)$$

where $dr$ is the differential increment in the distance, and $A$ is the surface area of the 3-dimensional (3D) sphere at a distance $r$. Hence,

$$dR = \rho \frac{dr}{4\pi r^2} \quad (6)$$

which implies

$$R = \frac{\rho}{4\pi} \int_0^r r^{-2} dr \quad (7)$$

Now, as shown in Fig. 8b, if we have a metal sphere/electrode with radius $r_e$ as the source (instead of a point charge), then equation (7) can be re-written to find the resistance up to an arbitrary point P as:

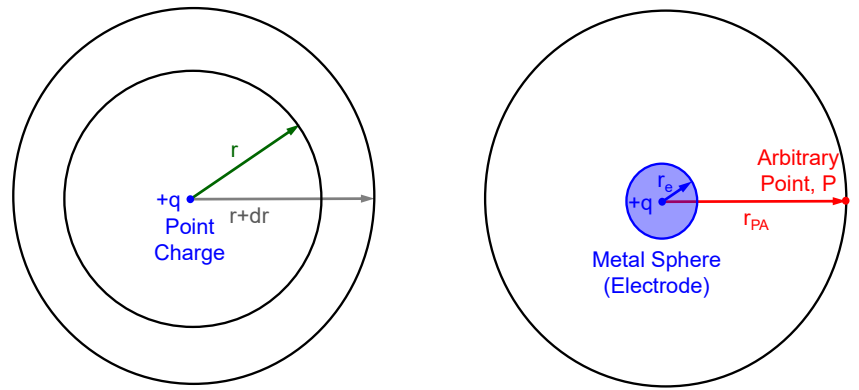

**a** With a Point Charge as a Source  **b** With a Metal Sphere as a Source

**Fig. 8 | Analysis of voltage developed at a distance from a Source. a,** Analysis with a point charge as a source. **b,** Analysis with a metal sphere (electrode) of radius $r_e$ as a source.

$$R = \frac{\rho}{4\pi} \int_{r_e}^{r_{PA}} r^{-2} dr \tag{8}$$

where P is an arbitrary point in 3D, at a distance of $r_{PA}$ from the center of the source. This results in:

$$R = \frac{\rho}{4\pi} \left[ -\frac{1}{r} \right]_{r_e}^{r_{PA}} = \frac{\rho}{4\pi} \left[ \frac{1}{r_e} - \frac{1}{r_{PA}} \right] \tag{9}$$

Therefore, the voltage drop across this resistance $R$ becomes

$$V_{drop} = IR = \frac{\rho I}{4\pi} \left[ \frac{1}{r_e} - \frac{1}{r_{dist}} \right] \tag{10}$$

Extending this idea for the dipole BP-QBC scenario (Fig. 7), and assuming that the voltage at point A is more than that of point B ($V_A > V_B$), the voltage at any arbitrary point P (in the 3D space) due to the charge at point A can now be written from equation (10) as:

$$V_{P,A} = V_A - \frac{\rho I}{4\pi} \left[ \frac{1}{r_{e,A}} - \frac{1}{r_{PA}} \right] \tag{11}$$

Similarly, the voltage at point P (in the 3D space) due to the charge at point B can be written from equation (10) as:

$$V_{P,B} = V_B + \frac{\rho I}{4\pi} \left[ \frac{1}{r_{e,B}} - \frac{1}{r_{PB}} \right] \tag{12}$$

Using the superposition theorem, we can write the voltage at point P as:

$$V_P = V_{P,A} + V_{P,B} = V_A + V_B + \frac{\rho I}{4\pi} \left[ \frac{1}{r_{e,B}} - \frac{1}{r_{e,A}} + \frac{1}{r_{PA}} - \frac{1}{r_{PB}} \right] \tag{13}$$

This is justified, since from equation (13), if we have a differential excitation ($V_A = -V_B$), and if the two electrodes at the UL TX are similar sized ($r_{e,A} = r_{e,B}$), then at the mid-point of A and B, $V_{P,mid} = 0$.

Also, it must be noted that at infinite distance, $r_{PA} \to \infty$, $r_{PB} \to \infty$, $V_P \to 0$ (boundary condition), which means

$$V_A + V_B + \frac{\rho I}{4\pi} \left[ \frac{1}{r_{e,B}} - \frac{1}{r_{e,A}} \right] = 0 \tag{14}$$

Putting the boundary condition (equation (14)) into equation (13), we get:

$$V_P = \frac{\rho I}{4\pi} \left[ \frac{1}{r_{PA}} - \frac{1}{r_{PB}} \right] \tag{15}$$

Equation (15) can now be utilized to find out the voltages at the outer surfaces of the UL TX electrodes (assuming finite conductivity of the metal electrodes) as shown in equation (16) and (17).

Outer surface potential ($V_{A'}$) of electrode with center point A:

$$V_{A'} = \frac{\rho I}{4\pi} \left[ \frac{1}{r_{A'A}} - \frac{1}{r_{A'B}} \right] = \frac{\rho I}{4\pi} \left[ \frac{1}{r_{e,A}} - \frac{1}{r_{AB} - r_{e,A}} \right] \tag{16}$$

Outer surface potential ($V_{B'}$) of electrode with center point B:

$$V_{B'} = \frac{\rho I}{4\pi} \left[ \frac{1}{r_{B'A}} - \frac{1}{r_{B'B}} \right] = \frac{\rho I}{4\pi} \left[ \frac{1}{r_{AB} - r_{e,B}} - \frac{1}{r_{e,B}} \right] \tag{17}$$

Therefore, the potential drop across the two UL TX electrodes is:

$$V_{A'B'} = V_{A'} - V_{B'} = \frac{\rho I}{4\pi}\left[\frac{1}{r_{e,A}} - \frac{1}{r_{AB}-r_{e,A}} - \frac{1}{r_{AB}-r_{e,B}} + \frac{1}{r_{e,B}}\right] \quad (18)$$

Similarly, equation (15) can be utilized to find out the voltages at the outer surfaces of the UL RX electrodes (assuming finite conductivity of the metal electrodes) as shown in equation (19) and (20).

Outer surface potential ($V_{C'}$) of electrode with center point C:

$$V_{C'} = \frac{\rho I}{4\pi}\left[\frac{1}{r_{C'A}} - \frac{1}{r_{C'B}}\right] = \frac{\rho I}{4\pi}\left[\frac{1}{r_{CA}-r_{e,C}} - \frac{1}{r_{CB}-r_{e,C}}\right] \quad (19)$$

Outer surface potential ($V_{D'}$) of electrode with center point D:

$$V_{D'} = \frac{\rho I}{4\pi}\left[\frac{1}{r_{D'A}} - \frac{1}{r_{D'B}}\right] = \frac{\rho I}{4\pi}\left[\frac{1}{r_{DA}-r_{e,D}} - \frac{1}{r_{DB}-r_{e,D}}\right] \quad (20)$$

Therefore, the potential (received voltage) across the two UL RX electrodes is:

$$V_{C'D'} = V_{C'} - V_{D'} = \frac{\rho I}{4\pi}\left[\frac{1}{r_{CA}-r_{e,C}} - \frac{1}{r_{CB}-r_{e,C}} - \frac{1}{r_{DA}-r_{e,D}} + \frac{1}{r_{DB}-r_{e,D}}\right] \quad (21)$$

However, at this point, we must note that the $\frac{\rho I}{4\pi}$ terms in equations (18) and (21) are different because of different amount of flux lines passing through the transmitting and receiving electrodes. From electrostatics, the $\frac{\rho I}{4\pi}$ terms in equations (18) and (21) can also be represented as $\frac{Q}{4\pi\varepsilon'}$, where $Q_{AB}$ and $Q_{CD}$ represent the effective EQS charges at the transmitting or receiving electrodes, respectively, to model the same inequality. The charges can be found from an electrostatic simulation in Ansys Maxwell.

The channel transfer function (TF), is therefore given by,

$$\text{TF} = \frac{V_{C'D'}}{V_{A'B'}} = Q_{CD}\left[\frac{1}{r_{CA}-r_{e,C}} - \frac{1}{r_{CB}-r_{e,C}} - \frac{1}{r_{DA}-r_{e,D}} + \frac{1}{r_{DB}-r_{e,D}}\right] \Big/ Q_{AB}\left[\frac{1}{r_{e,A}} - \frac{1}{r_{AB}-r_{e,A}} - \frac{1}{r_{AB}-r_{e,B}} + \frac{1}{r_{e,B}}\right] \quad (22)$$

which is same as equation (4).

As the channel length $= L$ ($r_{CA}$ or $r_{DB}$) increases, $V_{A'B'}$ remains constant. However, $V_{C'D'}$ will be proportional to $1/L^2$, indicating the signal transfer has a nature similar to 3D dipole coupling.

Fig. 4 in the main document compares the results of this analytical model with FEM simulation in HFSS. Although the analytical results largely follow the simulation results, there are some noticeable deviations (2-7 dB). These deviations arise because of the FEM simulation does not consider the channel as a homogeneous medium, and utilizes a detailed human head model consisting of realistic tissues (white matter + gray matter + CSF + Skull + Blood + Skin) which is taken from NEVA Electromagnetics[46]. Dielectric properties of the brain tissues have been obtained from the works of Gabriel et al.[21,22].

## Notes 2: Quantitative Comparison of Various Modalities of Data and Power transfer for Brain Implants

Fig 2c in the original manuscript represents a qualitative plot of some of the important considerations (tissue/skull absorption, form factor, transduction efficiency, robustness and transmitter depth) for neural implants using various modalities, including radio-frequency (RF), Inductive Coupling (IC), Ultrasound (US), Optical (OP), Magneto-Electric (ME) and BP-QBC. To provide more information about this qualitative figure, we have compiled the associated quantitative data in **Table 2**. The superscripts represent the corresponding additional references provided in the next two pages.

**Table 2 | Quantitative Comparison of available Data and Power Transfer Modalities:** Technologies such as Radio-Frequency (RF), Inductive Coupling (IC), Ultrasound (US), Optical (OP), Magneto-Electric (ME) and Bi-Phasic Quasistatic Brain Communication (BP-QBC) are considered.

| | RF | IC | US | OP | ME | BP-QBC |
|---|---|---|---|---|---|---|
| Brain Tissue Absorption (TA) | 1-2 W/kg, (~15dB/cm) at 900 MHz,[1,2] | ≈12 dB/cm, at 50-300 MHz,[7,8] | 10-12 dB/cm, at 10 MHz[13] | 10-15 dB/cm, around 650-950 nm,[18,19,20] | ≈ 6-7 dB/cm[26,27] at < 1 MHz | ≈ 10 dB/cm for up to 50 MHz[31] |
| Skull Absorption (SA) | 0.1-0.5 W/kg (≈0.7dB/mm) at 900 MHz,[1,2] | < 1 dB/mm at 50-300 MHz,[7,8] | 6-8 dB/mm at 10 MHz[13] | 1-1.5 dB/mm around 650-950 nm,[18,19,20] | ≈ 6-7 dB/cm[26,27] at < 1MHz | ≈ 10 dB/cm for up to 50 MHz[31] |
| Form Factor | 20 mm × 10 mm,[3] | 15 mm × 10mm,[7] | 4 mm × 0.75 mm,[14] | 0.25 mm × 0.4 mm,[21,22] | 5 mm × 4 mm,[28,29] | 4 mm × 2.5 mm[31] |
| Transduction Efficiency (TE) | 10% at 900 MHz,[4] | 1% at ≈100 MHz,[9] | 50%[15] | 0.1-10%[23,24] | > 0.01%[26,27,28,29] at < 1MHz | NA (Fully EQS)[31] |
| Robustness | Low, due to small aperture for RF coil | Low, due to coil alignment requirements | Highest | Med., due to light-induced parasitic short-ckt currents | High, due to less dependence on alignment than IC | High, less dependence on alignment |
| Transmitter Depth | 0.1-3 cm,[5,6] | 0.1-2 cm,[10,11,12] | 1-10 cm,[15,16,17] | ≈1 cm,[21,22,25] | Up to 15 cm,[30] | ≈ 6 cm,[31] |
| End-to-End System Loss (6 cm depth, no repeater/ Interrogator) | 10 dB (TE) + 75 dB (TA) + 5 dB (SA) ≈ 90 dB | 20 dB (TE) + 60 dB (TA) + 5 dB (SA) ≈ 85 dB | 3 dB (TE) + 60 dB (TA) + 50 dB (SA) ≈ 113 dB | 20 dB (TE) + 60 dB (TA) + 10 dB (SA) ≈ 90 dB | 40 dB (TE) + 35 dB (TA) + 5 dB (SA) ≈ 80 dB | 0 dB (TE) + 50 dB (TA) + 10 dB (SA) ≈ 60 dB |

*Note 1:* To calculate the end-to-end system loss, we assumed the following: thickness of skin and blood capillaries ≈ 2.5 mm, thickness of skull ≈ 7 mm, thickness of cerebro-spinal fluid ≈ 4 mm, and thickness of brain tissue (gray+white matter) ≈ 50 mm.

*Note 2:* Notice that IC has a TA of about 60 dB in the end-to-end system loss, while ME has only 20 dB TA, even though both of them traditionally use magnetic field coupling for signal transfer. This is because IC, in many cases, use a high frequency of operation just like RF to ensure that the coil sizes are not prohibitively large. This introduces significant amount of EM effects in the intended magnetic operation, and increases the tissue absorption. ME, on the other hand, operates in the magneto-quasistatic (MQS) range of frequencies (kHz to up to 10's of MHz), and hence offers lower tissue absorption.

*Note 3:* For OP systems, usually the absorption and scattering coefficients are provided, from which the dB loss is estimated. It needs to be kept in mind that different modalities use different figure of merit, but we have compared the dB loss for all of them.

*References used for Table 2:*